\documentclass{iau} 
\usepackage{graphicx}

\title[Understanding the Galaxy] 
{Understanding the Galaxy}

\author[Laurent Eyer]   
{Laurent Eyer$^1$
 }

\affiliation{$^1$Geneva Observatory,
University of Geneva, \\
Chemin des Maillettes 51,
CH-1290, Versoix, Switzerland \\ email: {\tt Laurent.Eyer@unige.ch}
}

\pubyear{2015}
\volume{xxx} 
\jname{Title of your IAU Symposium}
\editors{A.C. Editor, B.D. Editor \& C.E. Editor, eds.}
\begin{document}

\maketitle

\begin{abstract}
A general overview of the understanding of our Galaxy is presented following the
lines of its main structures: halo, disc, bulge/bar. This review is emphasising some
``Time Domain Astronomy'' contributions. On the one hand the distance and tangential
motion of the stars are essential to this understanding and are obtained through
multi-epoch surveys, on the other hand the chemistry of the stars, and their radial
velocity are also key elements to map Galactic \break (sub-)structures and unravel their
history and evolution. Contemporary surveys are revolutionising our view of the Milky Way
and galaxies in general. Among these, the Gaia mission excels by its precision astrometry
of 1.3 billion stars stretching through the Milky Way and beyond, providing the first 3D
view of a major part of the Milky Way. 
\end{abstract}

\firstsection 
\section{Introduction: Understanding our Galaxy}

Even the most simplifying satisfactory description of an object as complicated as our
Galaxy demands a high-dimensional parameter space covering distributions in 6D phase
space, age, and multi-dimensional chemistry of its stars and interstellar medium, as well
as the dark matter. These distributions are inherently complex, as there are several
components (disc, bulge/bar, halo) and at least the disc is persistently forming new
stellar generations. Furthermore, the Galaxy is not in equilibrium, showing transient
structures, e.g. transient spiral arms, current accretion events (e.g. the Sagittarius
dwarf) or bending waves and a disc warp.

Our ability to gather data, even though the current amount may look impressive, is still
limited and with these limited data there are some selection functions which can induce
strong changes depending on ``age, [Fe/H], distance" space, which strongly affects the
balance in samples between different populations; in addition statistical properties of
some fundamental parameters derived from some observables are very tricky, such as the
distance; finally we mention that some dynamical processes need to be accounted for before
we can decipher the Galactic past.

\begin{figure}[b]
\begin{center}
 \includegraphics[width=5.5in]{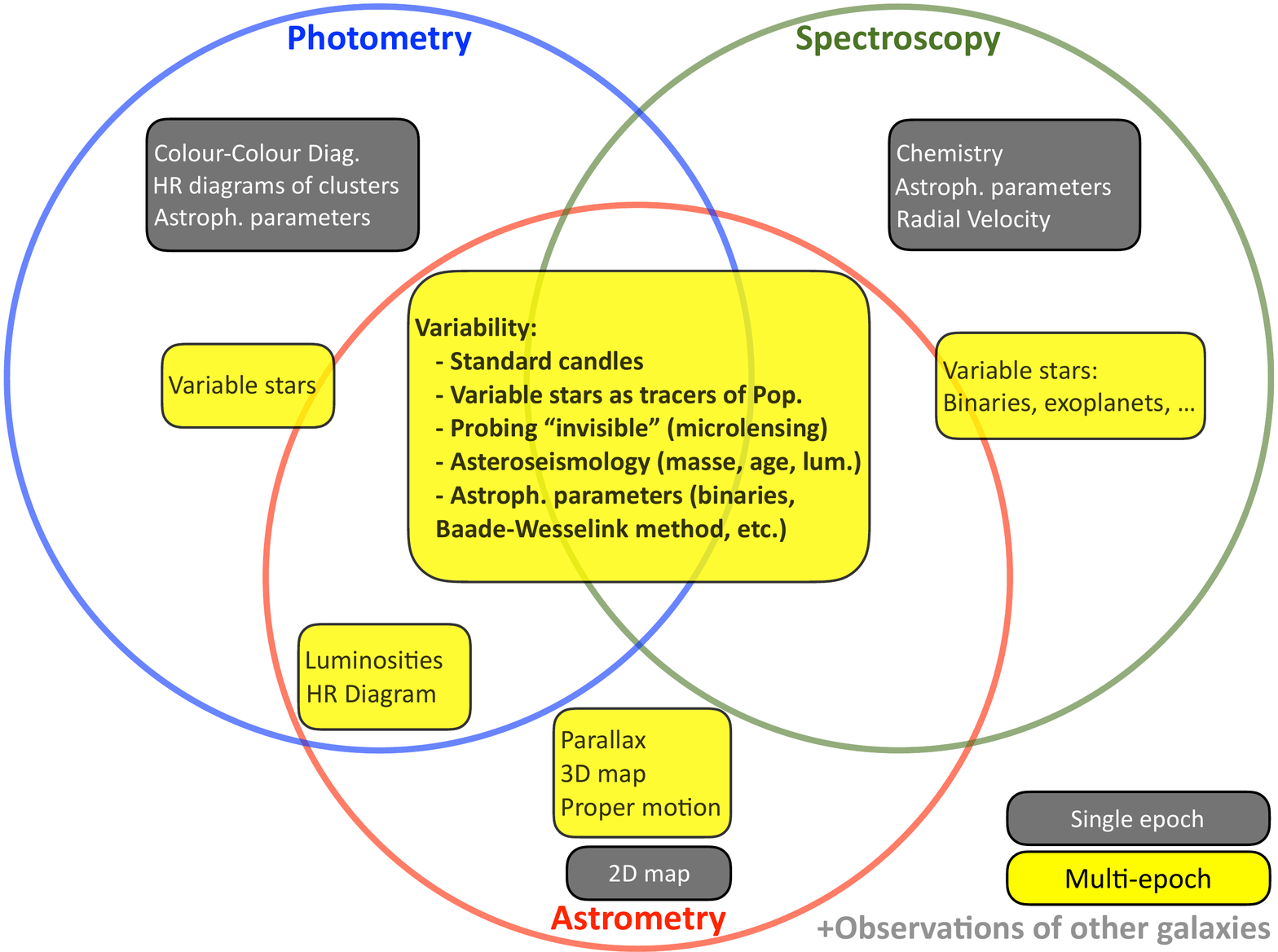} 
 \caption{Venn Diagramm of the three major observational techniques and the knowledge they
 contribute to. The grey boxes with white letters explicit what quantities can be derived
 with single measurements and the yellow boxes explicit what unique additional insight
 multi-epoch observations are providing.}
 \label{Fig1:ObsVennDiag}
\end{center}
\end{figure}

{\underline{\it The Observables}}. 
The knowledge in astronomy is mostly based on three main observational
techniques\footnote{There are additional techniques such as the detection of particles
(e.g. neutrinos), and gravitational waves.}: astrometry, photometry, spectroscopy. The
multi-epoch nature of these observables expands considerably the picture, as shown in
Fig.~\ref{Fig1:ObsVennDiag}. This figure highlights the central role of time domain
observations for nearly all fields of stellar and Galactic astronomy.

{\underline{\it Properties of stars from Time Domain Astronomy}}.
Stellar parallaxes and proper motions permit the production of 3D maps with tangential
motions, as well as the determination of absolute stellar magnitudes (though corrections
should be applied because of the interstellar extinction). There are however other
fundamental contributions of Time Domain astronomy: Some variable objects can be used as
tracers of specific populations (e.g. RR\,Lyrae stars for halo, Cepheids for young stellar
populations). Furthermore variable objects can bring knowledge on their fundamental
astrophysical parameters (e.g. thanks to the study of pulsating stars~-~asterosismology,
Baade Wesselink method~-~or binary stars). Some variability types can also be used to
establish the cosmic distance ladder (e.g. eclipsing binaries, Cepheids, RR\,Lyrae stars,
Supernovae).
%

{\underline{\it The Modelling Methods of the Galaxy}}.
The modelling of the Galaxy can be done by several broad approaches: (a) Jeans models:
This approach determines mass distributions from the observed kinematics by using Jeans
equation. (b) Made-to-Measure methods: such modelling could be achieved through N-body
simulations (which are computationally expensive), or Schwarzchild methods. In both
cases, the unicity of the model is not assured. i.e. several solutions may fit the data. 
With the Schwarzschild method, models are very difficult to achieve: there is a very large
parameter space, since one needs many orbits to describe a galaxy and therefore too few
orbits do not guarantee the right coverage of different orbit classes. In addition, the
stability of a solution is not assured. In addition, there is the need to make some
hypothesises on the dark matter distribution. (c) Action angle variables
(\cite{Binney2012}). Here, the system is considered in an axisymmetric equilibrium state.
The non-axysymmetry is treated as a perturbation of the model.

In order to reach a more complete description of our Galaxy, there is the additional need
to model its chemical evolution. 

{\underline{\it Some Examples of Simulations}}.
Some groups are simulating galaxies in isolation and are following the structure and the 
chemistry of the galaxies by doing self-consistent, N-body, chemo-dynamical modelling,
see \cite{RevazEtAl2016}. Other groups are simulating the formation of a galaxy in a
cosmological context, cf. for example the Fire-2
(\cite[Wetzel et al. 2016]{WetzelEtAl2016}), EAGLE
(\cite[Schaye et al. 2015]{SchayeEtAl2015}) or Illustris
(\cite[Vogelsberger et al. 2014]{VogelsbergerEtAL2014}) simulations.
These two approaches are touching in some ways on the debate ``nature versus nurture'', 
i.e. what is due to natural isolated evolution from physical laws versus what is due to
the perturbation of the environment or some initial conditions.

{\underline{\it The Data Flood}}.
On the observation facilities, we are in a very favorable period, technological and
computational developments, space missions are fully benefiting observational astronomy.
Such developments translate into a vigorous expansion of the three observational
techniques of Fig.~\ref{Fig1:ObsVennDiag}. This tremendous evolution is not only
quantitative but also qualitative: (1) in astrometry with Gaia; (2) in photometry with the
following projects Gaia,
LSST, 
Pan-STARRS, 
SkyMapper, 
OGLE,
SDSS,
Catalina,
PTF/ZTF,
Superwasp/NGTS,
VVV,
etc.
Photometric projects have also provided/will provide dense sampling for asterosismology:
Kepler,
CoRoT,
TESS,
Plato;
(3) in spectroscopy, there are also many surveys:
SDSS, SEGUE,
Gaia ESO Survey,
Gaia (RVS instrument),
APOGEE,
RAVE,
LAMOST,
Galah,
WEAVE,
4MOST,
etc.
\section{The Gaia mission: an exceptional Time Domain Survey}
Gaia is a space mission of the European Space agency which has started to impact many
fields of astronomy. In particular one of its primary science case is to study the 
composition, formation and evolution of the Galaxy (cf.
\cite[Perryman et al. 2001]{PerrymanEtAl2001}). Gaia fits well this goal, because its
instrumentation covers all aspects of Fig.~\ref{Fig1:ObsVennDiag}. Gaia is at heart a Time
domain mission. It is making an impressive jump in the astrometric precision. It is also
measuring/detecting tens of million of variable stars.

{\underline{\it The Gaia Data Releases}}. 
For the 5-year nominal mission, from 2014 to 2019, 4 data releases (DR) have been planned
(DR1: 2016, DR2: April 25 2018, DR3: 2020 and DR4: 2022-2023)\footnote{The mission
can be extended by at most 5 additional years due to the constrain of fuel
reserves. The ESA Science Programme Committee (SPC) has decided to extend the
mission up to end of 2020. Further extensions will be discussed at SPC in the general
context of extensions of all ESA missions.}. The second data release will contain a
catalogue of more than 1.3 billion stars with the 5 parameter astrometric solution. The
performance is magnitude dependent and will reach a parallax uncertainties of
0.04\,milliarcsecond for sources at G $<$ 15 and 0.7\,milliarcsecond at G = 20. This
second data release will also contain more than half a million of variable stars.

{\underline{\it Distances with Gaia}}. 
The distances derived from astrometric measurement are tricky. It is often better to work
in parallax space where in that case the assumption of symmetric distributions can be
often made, which is not the case for distances, see the article that will be published
with the second Gaia data release, Luri et al. (2018) in preparation.
\section{The Milky Way Global Structures in 2 Pages}
Several relatively recent reviews have been published on the Milky Way e.g. by
\cite{BlandHawthornGerhard2016} or chapters of Neill Reid within the 42nd Saas Fee
course book (\cite{SaasFee2015}). So here only a brief summary is written.

\subsection{The Galactic Halo} 
The stellar halo is the most extended structure of the Milky Way. The density of stars of
the halo follows a power law: $\rho = r^{-3}$, with a cusp at small radius and a cutoff at
large ones. The halo contains about 150 discovered Globular Clusters and a significant
part of the halo field stars stems from them (\cite{StarkenburgEtAl2009}).

The RR\,Lyrae stars reveal themselves extremely useful tools both for the description of
the Halo and Globular clusters. Thanks to RR\,Lyrae stars, \cite{Catelan2009} excludes
the possibility that the ``Galactic halo may have formed from the accretion of dwarf
galaxies resembling present-day Milky Way satellites''. \cite{SesarEtAl2017} was able to
trace in an unprecedented way the Sagittarius stream thanks to Pan-STARRS data. In fact
many streams and over densities have been observed in particular using SDSS data
(\cite{BelokurovEtAl2006}), some of these over-densities are due to accretion events, are
composed of disc stars (\cite{SheffieldEtAl2018}) or may emerge from vertical waves in the
outer disc (\cite{XuEtAl2015}). The interpretation of the streams is complicated as the
current observed location deviates from the actual trajectory of the accreted objects
(\cite{EyreBinney2011}). It is worth mentioning that it could be difficult to roll back
the history of mergers (\cite{JeanBaptisteEtAl2017}), because the integrals of motions are
not precisely conserved (e.g. further mergers, disc structure or simply dynamical
friction, etc.). Star diffuse through phase space in angular momentum due to stellar
migration, in the other actions due to heating, e.g. by spiral structure (mostly radial)
and molecular clouds (mostly vertical). Chemistry encodes the age and origin of each star
and thus helps us to reconstruct Galactic history and all this diffusion/redistribution.

The duality of the halo has been under debate, on its composition on one side and its
kinematical behaviour on the other, see \cite{BeersEtAl2012}, \cite{SchoenrichEtAl2011},
\cite{SchoenrichEtAl2014}. With the second Gaia data release, the halo kinematics will
undoubtedly be clarified.

The Milky Way halo was probed by several multi-epoch surveys to detect massive compact
halo objects (MACHOs), which was hypothetically forming the dark halo. Very few candidates
were found towards the Magellanic Clouds. However, microlensing events are common in the
direction of the bulge: About one variable star out of 500 is a microlensing event. These
multi-epoch surveys (MACHO, OGLE and EROS) brought light to many diverse results other
than constraining MACHOs (e.g. exoplanet detection through microlensing or transits,
discovery of many 100,000 variable stars and of new classes of variable stars, etc.).
Gaia will determine the distance of the nearest Globular Clusters and will detect many
variable stars such as RR\,Lyrae / SX\,Phoenicis stars.

A word on the dark halo: it might be close to spherical, see \cite{Read2014}, 
\cite{KuepperEtAl2015}, though this topic is debated. As written by \cite{Read2014}, 6D
phase-space information of Gaia will be transformative for this topic.

\subsection{The Galactic Disc}
The disc is made of many components, structures and substructures: field stars,
interstellar matter, giant molecular clouds, clusters, spiral arms, warps/waves.
Some of these structures are forming, others get transformed or even dissolve. Many
processes are at play, star formation, supernovae explosions, feedback, interaction with
satellites, accretion, etc. In all this complexity the density distribution of stars can
be described relatively simply by 2 exponentials profiles both in scale height and scale
length (height: 300 and 900\,pc and length: 2,600, 3,600\,pc resp., \cite{JuricEtAl2008}).
The stellar thin disc and its general properties can be explained by the properties of the
gaseous disc from which stars emerge with some heating mechanisms and angular momentum
changes. The thick component represents 30\% stellar surface density, it is composed by
older stars, having a lower averaged metallicity, a higher $\alpha$-element content and
larger random velocities. The origin of it has been under debate. It is established that
radial migration plays an important role (\cite{SellwoodBinney2002},
\cite{RoskarEtAl2008}, \cite{SchoenrichBinney2009}). However the full picture is not there
yet, because in this scenario a disc thickening of the inner parts is needed, so that
radial migration would transport the heat to outer disc regions to form the thick disc,
see \cite{AumerEtAl2017}.

Asteroseismology helps Galactic studies with age determinations for red giants branch
stars. Here, position in the colour-magnitude diagram becomes near-useless, as different
masses/ages converge onto the red giant branch, but the direct measurement from stellar
oscillations remedies this problem. An example of such study can be found in
\cite{CasagrandeEtAl2016}: combining Kepler data with Stroemgren photometry, the authors
found a vertical age gradient of 4\,Gyr/kpc, however there is large dispersion of ages at
all heights, and there is a flat age metallicity relation for disc stars.

On the side of the dark matter, the evidences of its presence in the form of dark disc
are weak, see \cite{Read2014}, \cite{SchutzEtAl2017}.

\subsection{The Galactic Inner Parts}

{\underline{\it The central black hole}}: Thanks to astrometric observations of stellar
orbits revolving around the black hole at the centre of our Galaxy, the mass of this
black hole has been constrained at the level of 5\% with more than 25 years of
observations, and is estimated to be 4.28 $10^6$ $M_{\odot}$ (\cite{GillessenEtAl2017}).
The black hole is also surrounded by a nuclear star cluster (\cite{SchoedelEtAl2009}).

{\underline{\it The nuclear disc}}: There were some published evidences of the
presence of a nuclear disc (\cite{CatchpoleEtAl1990}, \cite{LaunhardtEtAl2002}). More
recently a kinematic detection and characterisation of the inner disc with APOGEE spectra
were made by \cite{DebattistaEtAl2015} and \cite{SchoenrichEtAl2015}. In the latter study,
the radius of this inner disc is estimated to 150\,pc and its rotation velocity to
120\,km/s.

{\underline{\it The bulge/bar}}: The bulge/bar has been described by different groups
of stars. For example several variable star types were used to trace the bar: OSARG
(\cite{WrayEtAl2004}), Mira stars (\cite{CatchpoleEtAl2016}) or RR\,Lyrae stars
(\cite{PietrukowiczEtAl2015}). The bulge/bar can also be traced by red-clump stars, see
\cite{WeggEtal2015} and \cite{PortailEtAl2017}. In this last study Made-to-Measure models
were produced based on VVV, UKIDS, 2MASS, OGLE, BRAVA, ARGOS surveys and the bar pattern
speed was determined and stellar and dark matter
mass distributions were derived. From the above different studies, the bar orientation,
i.e. the angle with respect to the Sun-Galactic Centre line of sight, is still very
different from one study to the other.

Though there has been long standing arguments about the separated nature of the bulge and
bar and their origin, it seems that an in situ formation and evolution can explain
the present observations (\cite{DebattistaEtAl2017}).

\section{Conclusions}
Time Domain is at the root of our understanding of our Galaxy, though there is an
indispensable need of spectroscopy (for deriving the star chemistry and radial velocities)
to apprehend it fully. With the emergence of global multi-epoch surveys, and in
particular of Gaia and its astrometry, the understanding of our Galaxy will definitively
experience a significant leap forward.
~\\ 
~\\ 
{\underline{\it Acknowledgements}}. I am deeply indebted towards Dr Ralph Sch\"{o}nrich
for fascinating discussions during an invited stay at Oxford University, and also for
comments and corrections made on this written text. I would like also to thank Prof.
D.\,Pfenniger for fruitful discussions and comments on this text, Dr Y.\,Revaz and Dr
P.\,Eggenberger for interesting discussions.

\end{document}